\documentclass[12pt]{article}
\usepackage{times}
\usepackage{geometry}
\usepackage[utf8]{inputenc}
\usepackage{enumitem,amssymb}
\usepackage{ragged2e}
\usepackage{fancyhdr}
\usepackage{graphics,graphicx}
\usepackage{color}
\usepackage{xcolor}
\usepackage{lineno}
\usepackage{geometry}
\usepackage{wasysym}

\newlist{thematic}{itemize}{8}
\setlist[thematic]{label=$\square$}
\usepackage{pifont}
\usepackage{color}  
\usepackage{graphics,graphicx}   
\newcommand{\msun}{M$_{\odot}$}  

\begin{document}
\raggedright
\huge
Astro2020 Science White Paper \linebreak

Multi-Messenger Astrophysics Opportunities

\vspace{6pt}
with Stellar-Mass Binary Black Hole Mergers \linebreak
\normalsize

\vspace{12pt}
\noindent {Primary thematic area: \it \bf Multi-Messenger Astronomy and Astrophysics}

\vspace{4pt}
\noindent {Secondary thematic area: \it \bf Formation and Evolution of Compact Objects}

\vspace{12pt}
March 2019

\vspace{36pt}

\noindent Authors: \\
K. E. Saavik Ford (American Museum of Natural History and City University of New York)
Federico Fraschetti (University of Arizona)\\
Chris Fryer (Los Alamos National Laboratory)\\
Steven L. Liebling (Long Island University)\\
Rosalba Perna (Stony Brook University)\\
Peter Shawhan\footnote{Coordinating author. pshawhan@umd.edu, +1-301-405-1580} (University of Maryland and Joint Space-Science Institute)\\
P\'eter Veres (University of Alabama, Huntsville)\\
Bing Zhang (University of Nevada, Las Vegas)\\

\vspace{24pt}

\noindent Endorsers: \\
Vallia Antoniou (Texas Tech University and Harvard-Smithsonian Center for Astrophysics)\\
Dieter Hartmann (Clemson University)\\
Brian Humensky (Columbia University)\\
Szabolcs Marka (Columbia University)\\
Zsuzsa Marka (Columbia University)\\
Kohta Murase (Penn State University)\\
Marcos Santander (University of Alabama)\\
Colleen A. Wilson-Hodge (NASA Marshall Space Flight Center)\\
Stan Woosley (University of California, Santa Cruz)

\vspace{60pt}
\emph{This version has additional endorsers and a few small additions to the text relative to the version submitted to Astro2020.}

\vfill

\thispagestyle{empty}
\pagebreak
\pagenumbering{arabic}
\RaggedRight  


\section{Introduction}

The first gravitational-wave (GW) signal detected by LIGO, GW150914, was produced by the inspiral and merger of a pair of black holes (BHs) with masses of about 36 and 30 \msun\ \cite{2016PhRvL.116f1102A}, presumed to be the remnants of massive stars.  That discovery introduced us to a population of fairly heavy stellar-mass binary black hole (sBBH) systems, and a total of ten sBBH mergers have been confidently detected \cite {2016PhRvX...6d1015A,2018arXiv181112907T} in the first two observing runs of Advanced LIGO \cite{2015CQGra..32g4001L} and Advanced Virgo \cite{2015CQGra..32b4001A}.  Many more will be detected as the sensitivities of LIGO and Virgo improve and as more detectors (KAGRA in Japan \cite{2018arXiv181108079A}, and LIGO-India) join the network by the mid-2020s.  The enlarged network also will be able to localize events better, so that about half of the detected events will be localized to $\sim 10~\mathrm{deg}^2$ or better at 90\% confidence \cite{2018LRR....21....3A}.

\sloppypar Complementing the sBBH events, the binary neutron star (BNS) merger GW170817 was a watershed for multi-messenger astrophysics (MMA).  
It featured a strong GW signal followed closely by a gamma-ray burst \cite{GW170817-GRB}, a rich ``kilonova'' signature in the optical and infrared bands, and ultimately a multiwavelength afterglow which brightened over a period of months before fading (see \cite{GW170817-MMA}, \cite{2018arXiv180806617V} and many other papers).  Those emissions are understood to have come from disrupted neutron star matter that was ejected, some of which fell back to form an accretion disk and power a relativistic jet.
Of course, that source of matter is absent when two black holes merge, and so the conventional view of sBBH mergers is that there should not be enough matter present to produce a detectable electromagnetic (EM) transient.  However, there \emph{are} a number of mechanisms for producing such a counterpart, outlined below, and in fact, a weak gamma-ray transient signal was recorded by the {\it Fermi} Gamma-ray Burst Monitor (GBM) less than a second after GW150914 \cite{2016ApJ...826L...6C}.  The statistical significance of the association between the GBM event and GW150914 was less than 3 sigma, and no similar signal has been identified after other sBBH mergers so far, so it remains an intriguing but inconclusive hint of the possibility of gamma-ray emission from sBBH mergers \cite{2018ApJ...853L...9C}.

Our goals in this white paper are to 1) outline possible physical mechanisms for multi-messenger emission from sBBH mergers, and 2) describe the capabilities needed to either detect such emission or place substantive limits which can be translated into physical constraints.

\section{Models for EM emission from sBBH mergers}
The tentative $\gamma$-ray counterpart to GW150914 recorded by {\it Fermi}-GBM 
spurred several ideas and discussions on the possible presence of EM counterparts to sBBH mergers.  Since there is no material disrupted at merger, various ideas were put forward about astrophysical scenarios in which the merged BH would have some material to accrete from. Alternatively, an EM signal could be produced without accreting material, if the black holes are charged or through interactions with magnetic or exotic fields. 

\vspace{-1ex}
\subsection{Accretion models}
Among models invoking accretion, \cite{Loeb2016} proposed that the two sBHs which merged were produced within a dumbbell configuration in the core of a rapidly rotating star. That would provide abundant matter for accretion, but \cite{Woosley2016} argued that no single star of any mass and credible metallicity is likely to produce the GW signal recorded as GW150914.

\cite{Perna:2016jqh} proposed a model stemming from the evolution of two high-mass, low-metallicity and fast-rotating stars. The outer layers of the envelope of the second star to explode as a supernova remain bound and circularize at large radii in a fallback disk. With time, the disk cools and becomes neutral, suppressing the magneto-rotational instability, and hence the viscosity. The disk remains “long-lived dead” until tidal torques and shocks during the BBH pre-merger phase heat it up and re-ignite accretion, rapidly consuming the disk and providing energy to a possible outflow. Within this model, the time delay between the GW and a possible EM signal is given by the viscous time from the radius at which the viscous timescale is equal to the gravitational one, and it is found to be of a fraction of a second.
Additionally, thermal emission from a minidisk-powered wind could be seen as a fast optical transient with a duration from hours to days \cite{Murase2016}.

EM emission following a BBH merger could also be due to shock heating of a circumbinary disk left over from mass shed during the massive star evolution \cite{DeMink2017}. This would lead to medium-energy X-rays to infrared emission on timescales of a few hours following the GW event.

Another accretion-based model, put forward by \cite{Janiuk2017}, invokes a tight binary system made of a massive star and a BH which triggers the collapse of the star's nucleus, the formation of a second black hole, and eventually accretion of the remnant material from the star during the sBBH merger. It should be noted that, within the context of models involving BHs spiralling within the envelope of a massive star, energetic constraints become important so that the star does not become unbound prior to the merger \cite{Dai2017}. 

Alternatively, accreting material to fuel the BH engine post-merger could be provided by mass transfer in a hierarchical triple \cite{Chang2018}, or from
the large accretion disk of an active galactic nucleus (AGN) if the merging sBBH binary is located within it \cite{Bartos2017}.

Whatever the origin of material around the merging BHs, GRMHD simulations \cite{Khan2018} have shown the emergence of jets if some material is indeed present at merger. Therefore, the EM counterparts expected within most scenarios above would have characteristics similar to short GRBs (a prompt $\gamma$-ray signal followed by longer wavelength radiation over longer timescales), except for being less energetic (probably) if less mass is available for accretion.
Jets would also produce a flux of high-energy neutrinos.
(No TeV \cite{2016PhRvD..93l2010A} or MeV \cite{2018RAA....18..132Y} neutrino emission was detected from GW150914.)
However, since GWs are detected almost isotropically, while the high-energy emission is mostly concentrated within a narrow jet with an opening angle of order 10$^\circ$, the likelihood of detection, if the jet is relativistic, is higher at 
longer EM wavelengths, when the shock has slowed down. The brightness of the afterglow component would also be dependent on the density of the interstellar medium (see \cite{Perna2018} for a statistical study).

\vspace{-1ex}
\subsection{Charged black hole models}
The net electric charge Q on an astrophysical black hole is customarily assumed to be negligible because of the instability due to charge neutralization. However, that is not the full story.  For instance, a spinning BH in a uniform magnetic field is naturally charged \cite{Wald74}, although this charge is small.  In an astrophysical context, a charged pulsar (neutron star) will remain charged as long as it is spinning (\cite{michel82}; an opposite charge cannot directly enter the star to neutralize it, since the magnetosphere quickly self-adjusts), and if it collapses it naturally produces a Kerr-Newman black hole \cite{nathanail17}. Such a charged spinning BH possesses an external magnetosphere. How long the BH will retain this charge is a question subject to further study, but within the time frame of the simulation in \cite{nathanail17}, the amount of trapped charge remains constant at late epochs. This may be the most natural way to attain and retain substantial charges on astrophysical BHs.


A BH binary with at least one charged BH was proposed by \cite{Zhang:2016rli} to form a circular current loop during the inspiral, which results in a time-varying magnetic dipole moment. Along with the GW emission during the inspiral, the system undergoes an increase in the Poynting flux that generates a large electromotive force able to accelerate charged particles and to lead to a photon-pair plasma cascade. In analogy with radio pulsars, radio emission from the inner magnetosphere is produced via curvature radiation. The expanding plasma emits a burst in $\gamma$-rays via a possible number of processes already conceived for GRBs.  \cite{Zhang:2016rli} estimates that for a BH charge to mass ratio $q \simeq 10^{-5} - 10^{-4}$ (i.e., $Q \sim 10^{17} - 10^{18}$ C), a short GRB will be generated with a delay $\sim 1$ msec in the observer frame. If $q \simeq 10^{-9} - 10^{-8}$, a fast radio burst (FRB) can be produced. 

A similar scenario is studied in \cite{Liebling:2016orx} that solves numerically the Einstein equations coupled with the Maxwell equations for the final stage of the inspiral of two weakly-charged non-spinning BHs. The energy carried out as electromagnetic radiation across the entire spectrum per unit of time is consistent with the {\it Fermi}-GBM transient if the BHs hold an upper limit $Q \sim 10^{18}$ C (and $Q \sim 10^{12}$ C for an FRB); both values are comparable with \cite{Zhang:2016rli}. The time delay between the GW event and the observed EM luminosity peaks can be $< 1$ sec. 

In a different scenario, a net charge carried by two spinning collapsing objects was proposed by \cite{Fraschetti:2018}. If the two spins are anti-aligned, compression of the magnetic field, along with the collapsing low-density plasma within the current sheet separating the two magnetospheres, drives magnetic reconnection. Optically thick plasma flowing out from the reconnection layer dissipates magnetic energy into a relativistically hot pair plasma that gives rise to a short GRB at a distance $\sim 10^{14}$ cm from the merger, similar to the standard fireball scenario for GRBs. The luminosity of the {\it Fermi} transient entails a field of order $10^{12}$ gauss and a reconnection time-scale $\sim 10^{-2} - 10^{-3}$ sec for a reconnecting layer thickness $\ll 10^6$ cm, leading to a delay from the GW $< 1$ sec in the observer frame. The upper limit charge is in this case $Q \sim10^{19}$ C, within a factor $\sim$10 from the estimates above. All these models predict an approximate scaling of the EM luminosity as $Q^2$.

\vspace{-1ex}
\subsection{Field interaction models}

Accretion and BH charge represent the most conservative approaches to understanding a possible EM counterpart to a BBH GW event. However, less conventional explanations are worth exploring even if they ultimately only serve to constrain, and future detections may provide unambiguous signals that support some alternative explanation.

The Blandford-Znajek~(BZ) effect demonstrates that a BH spinning within a magnetic field sourced  by a circumbinary disk can generate
a jet-like region of collimated magnetic field~\cite{1977MNRAS.179..433B}. The magnetic field in this scenario must be sourced by an external current because a BH will promptly shed its own magnetic field (so that it has no ``hair''), and thus the BZ effect in binaries is generally applicable for supermassive BHs such
as might be found in merging galaxies that can easily provide material for such a circumbinary disk~\cite{Palenzuela:2010nf,Moesta:2011bn}.
For stellar-mass BBHs, the magnetic field could be provided by a circumbinary accretion disk assembled from local material.  A more exotic possibility is that the a BH has a monopole magnetic field, perhaps due to capturing a primordial cosmic monopole produced in the GUT phase transition of the Big Bang~\cite{Liebling:2016orx}.


Instead of a Maxwell EM field, perhaps other, exotic fields which couple to Maxwell surround BHs. Studies have considered
the stationary solutions and associated dynamics of BHs embedded within various fields from ultralight scalar to massive vector (see~\cite{2015IJMPD..2442014H,2018arXiv180605195B} for reviews). A binary black hole could provide an exceptional test or constraint of such physics.
Most of these studies consider only the GW signature~\cite{Baumann:2018vus,Ghosh:2018gaw}, but if any of these fields
do couple to the electromagnetic field, then an EM counterpart may be possible.
%
Exotic fields can also permit ultra-compact solutions---exotic compact objects (ECOs)---that may inspiral and merge like ordinary BHs, such as boson star binaries~\cite{Palenzuela:2017kcg} or other BH mimickers~\cite{Johnson-McDaniel:2018uvs,Helfer:2018vtq}.
Such a system may allow for a range of EM counterparts depending on the coupling of the ECOs to the Maxwell field.



\section{\mbox{What we can learn from MMA observations of sBBH mergers}}

The models described above present many scenarios for EM emission from a sBBH merger\linebreak detected by LIGO/Virgo/KAGRA.  We stand to learn about the astrophysics of these mergers and their environments in different ways, depending on the EM counterpart signature.  Multi-\linebreak messenger observations also enable a number of probes of cosmology and fundamental physics.

\vspace{-1ex}
\subsection{Stellar evolution and compact binary formation}

If an EM counterpart to a GW event is identified, the relative timing, spectrum and intensity of the EM emission will tell us about the emission mechanism and the characteristics of the sBBH system.
This complements what we learn from the GW data, such as the BH masses and the orientation of the binary orbit.
In accretion models, jet formation depends strongly on the location and amount of remnant mass around the binary (which, in some models, was outflow from one or both stars before collapsing to a BH) and how it is disturbed to activate accretion; thus we should be able to differentiate between models from observations (or the absence) of prompt and afterglow emission.
This can confirm or constrain the general picture of the GRB population \cite{Veres+16bbhgrb} as well as photospheric models \cite{Zhang+12phot,Veres+12phot}, and the fraction of sBBH mergers with observed beamed EM emissions will tell us about the beaming angle.
Finding a high-energy neutrino counterpart would further complete the picture of relativistic jet production.

The GW detector network in the mid-to-late 2020s will be able to reconstruct the location of a typical event to several square degrees or tens of square degrees \cite{2018LRR....21....3A}, sufficient to direct follow-up observations with X-ray, optical and radio telescopes if no \emph{prompt} EM counterpart were found with a precise location.
Finding an optical or radio counterpart will either place the event in a host galaxy or, if it was ejected from its host, tell us about the ``kicks'' given to the system when the BHs were formed.
From the metallicity and age of a host galaxy, we can study specific aspects of the binary populations producing binary black holes.  Population synthesis models have shown that both the merger rate and black hole mass distribution depend on metallicity and, as a result, host galaxy type \cite{belczynski10,dominik12}.  Comparing these models to observations can test our current understanding of metallicity effects on stellar evolution, with implications for a wide range of massive-star applications (supernovae, compact objects, Wolf-Rayet stars, etc.) \cite{1993ApJ...405..273W}, and interpretation of the BH mass distribution, including likely "mass gaps" below $\sim$4 \msun and above $\sim$50 \msun.
With age estimates, even more detailed comparisons can be made.
The location of the source \emph{within} the host galaxy also can tell us about the likely development of the progenitor system.

\vspace{-1ex}
\subsection{Cosmological measurements}


The EM counterpart to GW170817 enabled its host galaxy to be identified, and comparison of the distance determined from the GW signal with the galaxy's redshift yielded a measurement of the Hubble constant~\cite{Abbott:2017xzu}, later improved with a better determination of the binary orbit inclination from VLBI observations of the jet~\cite{2018arXiv180610596H}.
Future MMA observations of BBH mergers, along with neutron star binary mergers, will improve the precision of this method and may help resolve the present discrepancy between the two main Hubble constant measurement methods~\cite{DiValentino:2018jbh}, especially if EM counterparts are found for some fraction of sBBH mergers.  (In the absence of an EM transient counterpart, a statistical association is possible~\cite{Fishbach:2018gjp,2019arXiv190101540T}, but requires many more events to gain statistical power.)

 
\vspace{-1ex}
\subsection{Searches for charged black holes}

In the charged BH merger scenario, the EM luminosity essentially depends on the dimensionless charge \cite{Zhang:2016rli,Liebling:2016orx,Fraschetti:2018}, so non-detection of an EM counterpart to a sBBH merger can set an upper limit of the amount of charge possessed by the BHs. A large amount of charge would be required to explain the {\it Fermi}-GBM counterpart to GW150914; FRBs are a more plausible counterpart for charged BH mergers. The joint operation of wide-field FRB surveys such as CHIME \cite{chime} and GW detectors will soon enable interesting tests of this scenario. A joint detection would imply that at least one BH is the system is charged, and one may then use the FRB luminosity to estimate the amount of the BH charge.

\vspace{-1ex}
\subsection{Tests of general relativity (GR) and fundamental physics}

The onset time of a prompt transient EM counterpart can be compared to the GW event time to check whether the speed of GW propagation differs from the speed of light---predicted by some alternative theories of gravity---as was done with the BNS merger \cite{GW170817-GRB}.  Since sBBH mergers can be detected at much greater distances than neutron star binary mergers, sBBH events will measure or constrain any speed difference more precisely.  An EM counterpart which pins down the sky location more precisely than the GW data alone also will improve tests for extra GW polarization states or gravitational parity violation \cite{Yunes:2016jcc}.
Observations of BBH mergers serve to constrain modifications of GR in the very-strong-field and dynamic regime, complementing binary pulsar and solar-system tests.
Finally, EM observations can potentially detect, or else constrain, the presence of magnetic fields or the possibility of exotic fields modifying the BH spacetime solutions and coupling to EM wave emissions.

\vspace{-1ex}
\section{Implications for Astro2020: observing and modeling needs}

Multi-messenger observations of stellar-mass binary black hole mergers may seem speculative, but there are many plausible mechanisms leading to EM emissions, as we have reviewed above.  Some of these effectively produce a short GRB, while others lead to emission at longer wavelengths and over longer time scales.  To pursue these intriguing possibilities, we will need highly capable facilities observing in the 2020s and beyond, ranging from sensitive, all-sky monitors for gamma-ray bursts and X-ray bursts/afterglows to optical (including infrared) and radio observatories capable of very deep imaging.
A broad, robust observing campaign is needed to cover the range of possible luminosities, spectral content and light curve time scales.  
These will make it possible to address the science questions outlined in Section 3.

In addition, investments in modeling are needed to understand the EM signatures better, in order to optimize mission designs and to interpret any detected EM signal or to place constraints on astrophysical scenarios in the case of non-detection.

\pagebreak
\small
\bibliography{main}
\end{document}